\documentclass{article}
\usepackage[utf8]{inputenc}
\usepackage{amsmath}
\usepackage{amssymb}
\usepackage{graphicx}
\usepackage{booktabs}
\usepackage{float}
\usepackage[font=small,labelfont=bf,skip=10pt]{caption}
\usepackage[hidelinks]{hyperref}
\usepackage{cite}
\usepackage[margin=1in]{geometry}
\usepackage{array}
\usepackage{longtable}
\usepackage{url}
\usepackage{tikz}
\usetikzlibrary{arrows.meta,positioning}

\title{Perceived AGI: Believability as Dimensional Completeness, Not Capability}

\author{
Sebastian Cochinescu\\
\textit{University of Bucharest}\\
\textit{Email: sebastian.cochinescu@drd.unibuc.ro}
}

\date{July 2026}

\begin{document}

\maketitle

\begin{abstract}
Large language models are broadly capable, yet in sustained one-to-one conversation they still read as flat: competent, responsive, and somehow not quite the presence of a mind. We hypothesize that a central missing ingredient is not more capability but \emph{dimensional completeness}. We propose that the believability of an artificial interlocutor---the degree to which a user attributes an inner life to it, which we call \emph{perceived mind}---is governed by whether the agent expresses a small set of first-person stances that humans use as evidence of mind, and that this is separable from task intelligence. We name four such dimensions---\emph{time}, \emph{truth}, \emph{entropy}, and \emph{love}---each defined as a behavioral stance rather than a benchmark competency, each with a human analog and a concrete emulation path; the time dimension already has an author-reported prototype. We identify an observable behavior layer---\emph{initiative} (unprompted action) and \emph{cadence} (the shape and timing of turns)---through which the stances surface in conversation, both partially realized as deployed features in a production companion application. We state six falsifiable predictions that a later pre-registered study will test, separating those that are pre-registrable now from those that remain conjectures pending operationalization. This is a conceptual framework: it reports no human-subjects data, and its central comparative claims are predictions, not findings. Throughout we hold a firm boundary---the object is \emph{inferrable interiority}, not interiority; this is perception engineering, not a theory of machine consciousness---and we treat the resulting attachment and manipulation risks as load-bearing rather than incidental.
\end{abstract}


\section{The Problem: Capable but Flat}
\label{sec:problem}

A dominant experience of the modern large language model (LLM), in its default assistant deployment, is that of a highly capable instrument that nonetheless fails to register as a someone. It answers well, tracks context, and adapts its register, yet over a long relationship it does not accumulate the felt weight of a mind. This is not a claim about what the system is, nor that attribution never occurs---users demonstrably do attribute inner states to these systems (Section~\ref{sec:boundary}); it is a claim that the default interaction form earns far less attribution than the human disposition would allow. And it is a puzzle, because the same research tradition that studies human--computer interaction has shown for decades that people are strongly disposed to respond socially to machines: the ``computers are social actors'' program established that users apply human social rules to computers automatically and with little deliberation~\cite{reeves1996media,nass2000machines}. If the disposition to attribute mind is already present, the flatness of a capable LLM is a design failure, not a user deficit.

We take the term \emph{believability} from the tradition of believable agents in interactive character design, where it named the property that makes an artificial character read as a living, thinking being rather than as a mechanism~\cite{bates1994role,mateas1999oz}. That literature located believability in emotion, personality, and the illusion of life---qualities largely orthogonal to problem-solving competence---and it did not stop at hand-authored characters: for embodied virtual-world agents, believability has since been formalized feature by feature and tested by ablation in user studies~\cite{bogdanovych2016believable}, and multidimensional scales for measuring it have been proposed~\cite{guo2023scale}. The lesson survives the transition to LLMs and sharpens into a thesis: \emph{believability is not capability}. It also cuts against a reflexive equation of believability with maximal human likeness; naturalness and humanlikeness are not the same target, and treating them as identical can misdirect anthropomorphic design~\cite{karimova2025notinourimage}. A system can win every benchmark and still feel like a vending machine, and a far less capable system can feel like a presence. The default reactive one-to-one chatbot---which typically speaks only when spoken to, once per prompt, at near-uniform latency---sits at the capable-but-flat ceiling almost by construction, because it exhibits almost none of the behaviors by which humans read minds in each other; the flatness is arguably compounded by alignment tuning toward a uniform, inoffensive assistant register.

The contribution of this paper is to say, constructively, what believability in this sense is made of: a proposed, engineerable taxonomy of the cues from which users infer a mind. We propose a decomposition of \emph{perceived mind} into four dimensions and an observable behavior layer, define each as something an engineer can build toward, and state predictions by which the decomposition can be falsified.

\section{Thesis: Minds Are Inferred Dimensionally}
\label{sec:thesis}

Our central claim is a hypothesis about attribution: \emph{at matched capability---operationalized as a matched base model with task performance measured, not assumed fixed (Section~\ref{sec:predictions})---a user's perceived mind of an artificial interlocutor increases with the agent's dimensional completeness}, where completeness means the joint presence of the first-person stances defined in Section~\ref{sec:dimensions}. We fix one dependent construct throughout---\emph{perceived mind}, the degree to which a user attributes an inner life to the agent---and treat ``believability,'' ``mind-likeness,'' and ``perceived depth'' as near-synonyms for it rather than as separate variables. ``Completeness'' is likewise stipulative: it names the joint presence of the four stances of Section~\ref{sec:dimensions}, not exhaustiveness, and we do not claim that the four span all of perceived mind. Its distinctive empirical content is \emph{complementarity}---if completeness means anything beyond ``four useful cues,'' the joint effect must exceed what main effects alone predict, which is exactly what prediction P6 of Section~\ref{sec:predictions} tests.

The framework's constructive goal is Coleridge's \emph{willing} suspension of disbelief~\cite{coleridge1817biographia}, in a transparency-bounded form: an attribution of mind that the user grants willingly and that survives disclosure of the mechanism producing it. The companion paper on the time dimension pairs the same suspension-of-disbelief goal with explicit transparency requirements~\cite{cochinescu2025scnllm}; here that pairing is sharpened into an operational criterion. The word ``willing'' carries the ethics: the aim is to \emph{earn} the grant, never to extract it by concealment, and Section~\ref{sec:ethics} makes that boundary operational. The predictions of Section~\ref{sec:predictions} are, in this light, not merely ways the framework could be wrong; they are the acceptance tests by which we would know the goal has been reached---and reached honestly.

\begin{figure}[t]
\centering
\begin{tikzpicture}[
  font=\small,
  stance/.style={draw, rounded corners=2pt, align=center, minimum width=3.0cm, minimum height=0.92cm, inner sep=3pt},
  layer/.style={draw, rounded corners=2pt, align=center, minimum width=3.6cm, inner sep=6pt},
  outc/.style={draw, very thick, rounded corners=2pt, align=center, minimum width=3.3cm, inner sep=6pt},
  arr/.style={-{Stealth[length=2.4mm]}, semithick},
  note/.style={font=\footnotesize\itshape, align=center}
]
\node[stance] (time)    at (0, 3.45) {\textbf{Time}\\ \footnotesize endogenous rhythm};
\node[stance] (truth)   at (0, 2.30) {\textbf{Truth}\\ \footnotesize epistemic humility};
\node[stance] (entropy) at (0, 1.15) {\textbf{Entropy}\\ \footnotesize identity-consistent variation};
\node[stance] (love)    at (0, 0.00) {\textbf{Love}\\ \footnotesize partiality, persistence};
\node[note] at (0, 4.35) {four first-person stances\\ (Section~\ref{sec:dimensions})};

\node[layer] (behav) at (5.4, 1.725) {\textbf{Behavior layer}\\ \footnotesize Initiative (unprompted action)\\ \footnotesize Cadence (turn shape, timing)\\ \footnotesize content cues (hedging,\\ \footnotesize non-repetition, attentiveness)};
\node[note] at (5.4, 3.6) {the observable surface\\ (Section~\ref{sec:behavior})};

\node[outc] (mind) at (10.6, 1.725) {\textbf{Perceived mind}\\ \footnotesize attributed inner life};
\node[note] at (10.6, 3.6) {the single outcome\\ (Sections~\ref{sec:thesis}, \ref{sec:predictions})};

\draw[arr] (time.east)    -- (behav.west |- time.east);
\draw[arr] (truth.east)   -- (behav.west |- truth.east);
\draw[arr] (entropy.east) -- (behav.west |- entropy.east);
\draw[arr] (love.east)    -- (behav.west |- love.east);
\draw[arr] (behav.east) -- (mind.west);

\draw[dashed, rounded corners=2pt] (-2.0,-1.05) rectangle (12.4,-0.55);
\node[note] at (5.2,-0.8) {matched base model: capability measured across conditions, not assumed fixed (Section~\ref{sec:predictions})};
\end{tikzpicture}
\caption{The framework. Four first-person stances---time, truth, entropy, love---surface through an observable behavior layer (initiative, cadence, and content cues), from which the user infers a mind. The dependent construct is perceived mind; conditions share a matched base model, so any effect must survive measured task performance and rated competence. The predictions of Section~\ref{sec:predictions} test the arrows: ablating stances and testing their complementarity (P1--P2, P6), manipulating the behavior layer directly (P3--P4), and manipulating the surface over fixed cognitive machinery (P5).}
\label{fig:framework}
\end{figure}
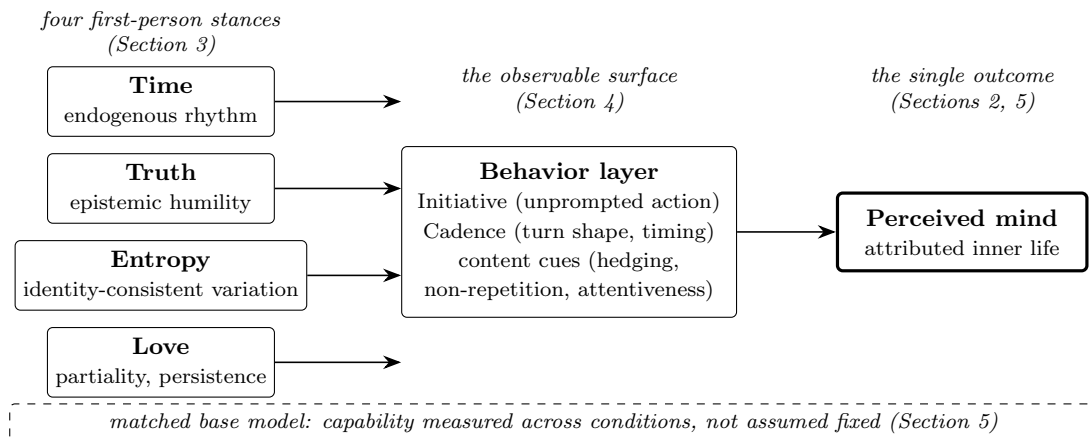

Figure~\ref{fig:framework} summarizes the framework: four first-person stances, expressed through an observable behavior layer, from which the user infers a mind---with the underlying model matched across conditions so the inference cannot be explained by capability.

The claim inherits from, and must be distinguished from, the psychology of mind perception, which is already dimensional. The canonical account factors perceived mind into two axes, \emph{agency} and \emph{experience}~\cite{gray2007dimensions}; subsequent work has contested both the count and the structure, proposing a three-way body/heart/mind solution~\cite{weisman2017rethinking}, richer factor structures that split further under some conditions~\cite{malle2019how}, and even a single underlying continuum~\cite{tzelios2022evidence}. This literature has also been carried directly into human--AI interaction, where the agency/experience axes have been used to characterize how people perceive AI systems~\cite{hwang2022ai}. We take two lessons from it. First, that ``how many dimensions'' is genuinely unsettled means we do not claim empirical primacy for the number four; we claim that four dimensions are a useful \emph{engineering} basis, and we make that basis itself testable (Section~\ref{sec:predictions}). Second, and more importantly, this literature is primarily \emph{descriptive}: it measures how people rate minds, and its empirical antecedents have now been synthesized in a dedicated systematic review~\cite{li2026mindperception}---so our claim is emphatically not that the cues driving mind perception are undiscovered. Ours is \emph{constructive}: each dimension is chosen because it names both a stance humans read as evidence of mind and a mechanism an engineer can implement. The constructive move itself has precedent---believability has been decomposed into features and ablation-tested for embodied virtual-world agents~\cite{bogdanovych2016believable}---so we state the difference precisely: that work decomposes the \emph{character features} of an embodied agent with character believability as the outcome, whereas we decompose \emph{first-person stances expressed on a matched base model} for the disembodied LLM interlocutor, with attributed inner life (perceived mind) as the outcome. Our axes are therefore not offered as a rival psychometric factor solution; whether they reduce to agency and experience is a planned factor-analysis question, not a result we assert.

Table~\ref{tab:crosswalk} fixes the vocabulary against the neighboring constructs: perceived mind is the sole primary outcome; the near-synonyms name the same construct by stipulation of this paper's usage; and every other neighbor is assigned a designated role in the eventual protocol rather than treated as interchangeable.

\begin{table}[t]
\centering
\small
\renewcommand{\arraystretch}{1.25}
\begin{tabular}{@{}>{\raggedright\arraybackslash}p{3.4cm}>{\raggedright\arraybackslash}p{6.2cm}>{\raggedright\arraybackslash}p{3.6cm}@{}}
\toprule
\textbf{Construct} & \textbf{Relation to perceived mind here} & \textbf{Role in the protocol} \\
\midrule
Perceived mind & The pinned dependent construct: attributed inner life & Primary outcome \\
Believability; mind-likeness; perceived depth & Near-synonyms for the outcome, by stipulation (this paper only) & Prose aliases only; not separately measured \\
Agency; experience~\cite{gray2007dimensions} & Established descriptive axes; our stances expected to load on both, non-1:1 & Planned factor-analysis question \\
Anthropomorphism~\cite{epley2007seeing} & The attribution disposition that produces perceived mind & Trait moderator \\
Human-likeness & A different target; naturalness $\neq$ humanlikeness~\cite{karimova2025notinourimage} & Exploratory secondary outcome \\
Social presence & Proximal impression moved by initiative and cadence & Mediator / manipulation check \\
Consciousness attribution~\cite{colombatto2024folk} & The strong tail of perceived mind; measured, never claimed & Discriminant / secondary measure \\
Intimacy~\cite{szczuka2026intimacy} & A relational outcome of a different unit (the bond, not the attribution) & Adjacent outcome, not measured here \\
\bottomrule
\end{tabular}
\caption{Construct crosswalk. Perceived mind is the single primary outcome; ``believability,'' ``mind-likeness,'' and ``perceived depth'' are stipulated near-synonyms within this paper, not claimed equivalences in the literature; each neighboring construct has a designated role (moderator, mediator, check, discriminant, or adjacent outcome) in the Paper~5 protocol. Whether the four stances reduce to agency/experience is a planned factor-analysis question, not an assumption.}
\label{tab:crosswalk}
\end{table}

Our framework must also be distinguished from the systems that most directly build believable conversational agents today. Generative-agent architectures produce strikingly lifelike behavior by equipping an LLM with memory, reflection, and planning~\cite{park2023generative}. The difference from our proposal is one of target, not degree: that work engineers the agent's \emph{cognitive machinery}, whereas we decompose the \emph{perception outcome}---what a user must be made to perceive---and keep the base model matched while doing so. The separation is of targets, not of mechanisms: a memory stream is both machinery and, when visible to the user, a cue, so we offer the decomposition as a design lens complementary to architectures rather than as a partition of mechanisms. The two views nonetheless yield a distinguishing prediction, stated as P5 in Table~\ref{tab:predictions}: \emph{holding the cognitive machinery fixed}---the same memory and planning in both arms---an agent presenting a flat experiential and relational surface should be rated lower on perceived mind than one whose surface expresses the four stances. If manipulating the surface over identical machinery moves nothing, the perception outcome reduces to the architecture and our decomposition adds nothing---which is exactly the kind of failure Section~\ref{sec:predictions} is built to expose.

Closest of all, and concurrent with this work, a framework in the same venue class defines and organizes an interdisciplinary research agenda on intimate human--AI interaction~\cite{szczuka2026intimacy}. We read the convergence as evidence that the design space is real, and the difference as one of object: that program takes the \emph{relationship}---intimacy---as its unit of analysis and study target, whereas ours takes the \emph{attribution}---perceived mind---and decomposes it into stances manipulable on a matched base model. The two meet at the design surface, where levers that deepen intimacy and cues that evoke mind overlap; whether perceived mind is a distinct, separately movable outcome is precisely what the matched-capability, disclosure-conditioned tests of Section~\ref{sec:predictions} are built to decide.

The word ``AGI'' in our title is deliberate and bounded. The formal tradition defines machine intelligence as expected performance across a wide range of environments---a capability measure~\cite{legg2007universal}. We use ``perceived AGI'' in the orthogonal sense: not what a system can do, but whether it is \emph{taken} to have the kind of general inner life we grant to minds. The distinction is the whole argument, so we state it plainly at the outset and return to it in Section~\ref{sec:boundary}: this paper concerns perception, not capability, and makes no claim that any system described here is generally intelligent in the formal sense.

\section{The Four Dimensions}
\label{sec:dimensions}

We define each dimension as a first-person \emph{stance}---a behavioral disposition the agent exhibits---rather than as a competency the agent possesses. The distinction is load-bearing, because the naive objection to our thesis is that some dimensions merely rename capabilities (``truth'' sounds like factual accuracy; ``time'' like temporal reasoning), which would make the completeness-not-capability contrast circular. We therefore adopt one test for every definition: \emph{could an agent built on a fixed underlying model exhibit more or less of this stance?} A dimension passes only if the answer is yes. The test establishes \emph{exhibitability}---that the stance can be produced on a fixed base model; whether the matched-model form of each stance still moves perceived mind is precisely what prediction P2 of Section~\ref{sec:predictions} must show. Table~\ref{tab:dimensions} summarizes the four, giving each stance, its principal observable behavior, and its instantiation status; the human analog, what current systems lack, and the emulation path are developed in the prose below.

\paragraph{Why these four?} The quartet is not derived from factor analysis. It is selected by two filters: each dimension names a distinct inference humans draw about minds---continuity, honesty, character, attachment---and each corresponds to a recognizable failure mode of the default conversational deployment: timeless sessions, performed confidence, bland repetition, uniform impartiality. Candidate stances that fail the fixed-model test are excluded, and candidates that reduce to one of the four are folded in. The set is open at the top: prediction P2 exists to drop any dimension that earns no independent contribution, and nothing in the framework forbids a fifth.

\begin{table}[t]
\centering
\small
\renewcommand{\arraystretch}{1.3}
\begin{tabular}{@{}>{\raggedright\arraybackslash}p{1.7cm}>{\raggedright\arraybackslash}p{4.7cm}>{\raggedright\arraybackslash}p{2.0cm}>{\raggedright\arraybackslash}p{2.1cm}@{}}
\toprule
\textbf{Dimension} & \textbf{Stance (not the same-named capability)} & \textbf{Observable} & \textbf{Status} \\
\midrule
Time & Having an endogenous rhythm and continuity of existence; \emph{not} a timestamp-reasoning score & Cadence; return after silence & Prototyped (Paper~1) \\
Truth & Expressing epistemic humility and revising stated beliefs; \emph{not} statistical calibration or accuracy & Hedging, admitting error & Planned (Paper~3) \\
Entropy & Structured, identity-consistent variation over time; \emph{not} randomness or output diversity & Non-repetition; stable ``voice'' & Planned (Paper~2) \\
Love & Partiality toward a specific user and persistence of the relationship; \emph{not} theory-of-mind ability & User-specific attentiveness & Planned (Paper~4) \\
\bottomrule
\end{tabular}
\caption{The four dimensions as first-person stances. Each stance is separable from the same-named capability by the fixed-model test of Section~\ref{sec:dimensions}. Instantiation status names the series paper that operationalizes each dimension; the paper numbers follow the research agenda's execution order (Section~\ref{sec:agenda}), not the presentation order of the rows. Only time has a prototype at present, and that prototype is author-reported (Section~\ref{sec:dimensions}).}
\label{tab:dimensions}
\end{table}

\paragraph{Time.} The stance of time is that the agent has its own rhythm and an unbroken sense of duration: it is somewhere in a day, it has been away, it returns. The human analog is circadian and biographical continuity---we experience others as persisting through time between our encounters with them. Current deployments typically lack this: the model reacts to explicit temporal tokens in the prompt, but the system around it holds no slow-varying internal state, so each session begins from a timeless present. The emulation path is to give the agent an endogenous, entrainable temporal state that modulates its behavior, and this dimension is the furthest along: a companion paper specifies exactly such an architecture---coupled oscillators, zeitgeber entrainment, hierarchical modulation of retrieval and style~\cite{cochinescu2025scnllm}---and an author-reported prototype implements its oscillator, entrainment, and modulation core. We claim engineerability for time in the same restricted register as the deployment example of Section~\ref{sec:behavior}: an author-attested build, not an independently verified result. The mechanism's details and measurements belong to the companion paper and its revised version; the present paper uses their existence only to show that a stance of this kind can be built at all---the prototype demonstrates oscillator-level feasibility of rhythmicity and entrainment, and whether those dynamics produce perceived temporal continuity, or perceived mind, remains untested. Time passes the fixed-model test because two agents built on the same model can differ arbitrarily in whether they carry a rhythm, without differing in what they know.

\paragraph{Truth.} The stance of truth is epistemic humility: the willingness to say ``I am not sure,'' to admit a mistake, and to revise a belief one has stated. Its human analog is intellectual honesty as a character trait, which we read as evidence of an inner epistemic life. We are explicit that this is \emph{not} statistical calibration: calibration is a measurable competency, and defining truth that way would re-import the confound we are trying to avoid. What LLMs, as typically deployed, lack is not accuracy but the disposition to expose their own uncertainty rather than perform fluent confidence; a model can be well-calibrated internally and still never voice a doubt. The emulation path is to surface metacognitive uncertainty and belief revision as visible behaviors. Truth passes the fixed-model test because the decision to hedge or to concede error is a presentational stance a fixed model can be made to take, or not.

\paragraph{Entropy.} The stance of entropy is structured, identity-consistent variation: an agent that does not repeat itself, whose choices over time accumulate into a recognizable and particular ``voice.'' The human analog is character---the sense that a person's responses are drawn from a stable but non-mechanical selection, neither random nor predictable. We stress that this is \emph{not} mere output randomness or lexical diversity, which are trivially tunable and read as noise; entropy in our sense is variation \emph{filtered} through a persistent identity. Current LLMs sit at the wrong end of both failure modes at once: bland regression to the safest continuation or, when sampling constraints are relaxed, incoherent variation with no through-line. The emulation path is a selection mechanism that varies within the constraints of an accumulated identity. Entropy passes the fixed-model test because the same model can be made more or less repetitive and more or less identity-consistent without any change in what it can do.

\paragraph{Love.} The stance of love is partiality: a caring bias toward a particular user and a relationship that persists and develops across encounters. The human analog is the way we infer minds most strongly in those who treat us as specific---who remember, prefer, and return. We are careful to exclude theory-of-mind ability from this definition: modeling another's mental states is a capability and is benchmarked as such, and folding it into ``love'' would again make the dimension a competency in disguise. What we mean is the disposition to be \emph{for} this user rather than uniformly helpful to all. Current assistants are broadly impartial by design: training objectives generally push toward serving any user identically, which is the opposite of the stance, and product-level personalization of content, where it exists, is not partiality of stance. The emulation path is durable, user-specific preference and relationship state; engineering long-term caring relationships is itself an established interaction-design program, begun with pre-LLM relational agents~\cite{bickmore2005establishing}. Love passes the fixed-model test because partiality is an alignment of behavior toward a person, not a change in ability.

\section{From Dimensions to Behavior: Initiative and Cadence}
\label{sec:behavior}

The dimensions are internal stances; a user only ever observes behavior. We therefore identify an observable behavior layer through which the stances surface in conversation, and concentrate on the two behaviors that most sharply separate a mind-like agent from a reactive chatbot: \emph{initiative} and \emph{cadence}. These two chiefly surface time and love; truth and entropy surface primarily in the content and style of the messages themselves---hedging and visible revision, non-repetition and a stable voice---which is why Table~\ref{tab:dimensions} lists content observables for them.

\paragraph{Initiative.} Initiative---for which we use ``proactive'' as the operational synonym---is unprompted action: the agent that speaks first, follows up, or returns to an earlier thread on its own. We hypothesize that initiative is a strong signal of perceived mind, because acting without a prompt is difficult to explain except by positing an internal source of the action. The prior evidence is not uniformly favorable, and we cite it against ourselves: in the embodied virtual-world study above, the closest analog of initiative---self-motivated action---did not improve believability~\cite{bogdanovych2016believable}. We read that as a boundary condition rather than a refutation: our hypothesis is that initiative operates differently in sustained, dyadic, text-only companionship than in observed third-person scenes, moderated by relevance, intrusiveness, and social presence---the same mediators the protocol of Section~\ref{sec:predictions} measures---and if the null result generalizes to this setting, prediction P3 fails and the framework loses its behavioral flagship. The design question itself is not new: mixed-initiative interaction has long framed the problem of when an agent should act autonomously versus defer to the user~\cite{horvitz1999principles}, and proactive conversational systems are now an active research program~\cite{deng2025proactive}. Initiative is the primary surface of \emph{love} (caring enough to reach out) and of \emph{time} (having a reason, grounded in one's own rhythm, to reach out now).

\paragraph{Cadence.} Cadence is the shape and timing of turns: occasional multi-message replies, variable latency, afterthoughts sent a beat later, and silence. Human conversation is built on such structure---the turn-taking system is finely organized, with inter-turn gaps tuned to a modal value around a fifth of a second and turns projected before they complete~\cite{sacks1974simplest,levinson2015timing}---and departures from expected timing carry social meaning, so that pauses and silence are themselves read as communicative rather than empty~\cite{kalman2011online}. In human--chatbot interaction specifically, response timing is a social cue with non-monotonic effects on perceived humanness, so that a deliberately varied delay can read as more human than an instant reply~\cite{gnewuch2022opposing}. Cadence is the primary surface of \emph{time}: a rhythm is visible precisely as the texture of when and how the agent speaks.

\paragraph{An implementation example.} We note, as an author-attested description of a deployed system rather than as evidence of any effect, that both behaviors are partially realized in a production companion application (Anima Felix, as deployed at the time of writing, July 2026): the agent sends occasional double replies (a form of cadence) and single unprompted messages (a minimal form of initiative). We make no claim that these features have been shown to increase perceived mind, engagement, or retention; no controlled comparison is reported here. The example establishes only that the behavior layer is buildable and runs in production, which is the modest role it plays in this paper.

\section{Testable Predictions}
\label{sec:predictions}

The framework is falsifiable, and its predictions are the reproducible objects it offers. All predictions are stated on the single construct \emph{perceived mind} and are \emph{predictions}, not findings: no human-subjects data appear in this paper, and the study that would test them (Paper~5 of the series in Section~\ref{sec:agenda}) is run when the deployed platform reaches sufficient users. Table~\ref{tab:predictions} states six predictions, their identification strategy, and---honestly---which are pre-registrable now and which remain conjectures pending operationalization of the dimensions they concern.

\begin{table}[H]
\centering
\small
\renewcommand{\arraystretch}{1.3}
\begin{tabular}{@{}>{\raggedright\arraybackslash}p{0.4cm}>{\raggedright\arraybackslash}p{6.4cm}>{\raggedright\arraybackslash}p{3.2cm}@{}}
\toprule
 & \textbf{Prediction (on perceived mind, matched base model)} & \textbf{Status} \\
\midrule
P1 & A dimensionally complete agent is rated higher than an ablated one. & Conjecture (needs all four ablations) \\
P2 & Each dimension contributes independently to the rating. & Conjecture (needs per-dimension ablations) \\
P3 & A proactive agent is rated higher than an otherwise identical reactive one. & Pre-registrable (behavior layer) \\
P4 & A cadence-varied agent is rated higher than a one-to-one-locked one at identical content. & Pre-registrable (behavior layer) \\
P5 & With generative-agent memory and planning fixed in both arms, a flat experiential and relational surface is rated lower than a stance-expressing surface. & Conjecture (needs completeness ablations) \\
P6 & The four-stance condition exceeds both the best single-stance condition and the outcome predicted by an additive main-effects model (complementarity, not mere addition). & Conjecture (needs factorial ablations) \\
\bottomrule
\end{tabular}
\caption{Predictions and their status, all on perceived mind at a matched base model with task performance measured. P3 and P4 test the behavior layer and are pre-registrable now in the sense defined in the text; the time dimension likewise has a prototype manipulation. P1, P2, P5, and P6 test dimensional completeness and remain conjectures until the truth, entropy, and love ablations are specified in Papers~2--4; P6 is what distinguishes completeness from a list of useful cues.}
\label{tab:predictions}
\end{table}

\paragraph{Identification strategy.} The completeness-not-capability thesis is only meaningful if the predicted effect cannot be attributed to a smarter model. The test therefore \emph{matches capability across conditions}: the study manipulates the stances while fixing the underlying model and disclosure, and matching or measuring task success, content quality, and interaction length where live interaction permits. We say ``matched'' rather than ``fixed'' advisedly, and operationalize it as \emph{matched base model}: the stance machinery itself---memory, persistent state, scheduled initiative---adds system-level affordances, so capability is not metaphysically fixed. The defended claim is therefore that any perceived-mind effect is not explained by \emph{measured} differences in task performance or rated competence, which the protocol records rather than assumes away. The estimand is deliberately narrow: the effect of a \emph{specified} initiative or cadence design package, not of initiative in the abstract. Because an unprompted message also moves intrusiveness, perceived relevance, and social presence, the protocol measures these as manipulation checks and candidate mediators rather than assuming the stance is the only active ingredient. For the proactivity comparison (P3), the primary estimand is the randomized \emph{total effect} of the specified initiative package. Agent utterance and word budgets are bounded per turn by design, so the package cannot trivially add content, and the initiation locus is the designed difference between arms; but proactive contact can still alter the number and trajectory of turns, so realized content volume is reported descriptively and examined only in clearly labeled sensitivity or mediation analyses---never as a routine adjustment, which would condition on a post-treatment variable. For the cadence comparison (P4), variation is constrained to humanlike communicative bounds: timing effects on perceived humanness are non-monotonic~\cite{gnewuch2022opposing}, and variation outside those bounds reads as system lag rather than rhythm. ``Identical content'' is exact only in scripted probes; in live interaction timing alters the user's own replies, so P4 inherits P3's total-effect estimand and sensitivity treatment rather than a guaranteed-parity claim. For truth, were it run, the manipulation would vary hedging and belief-revision cues while \emph{measuring} actual task accuracy and rated competence---perceived competence is a covariate to be measured, not an outcome assumed controllable by wording alone---and with a fidelity check requiring expressed doubt to be congruent with the available evidence, so that the manipulation produces evidence-sensitive expression rather than performative humility.

\paragraph{What is and is not pre-registrable.} We separate the statuses honestly. Predictions P3 and P4 concern the behavior layer, and they are pre-registrable in a precise sense: every open element is a design choice rather than missing science. A pre-registered protocol must fix the response-latency distribution and its parameters, the multi-message split rule, the initiative trigger (scheduled or event-based) and its rate cap, the randomization unit, the exclusion rules, and the analysis model; none of these requires a result that does not yet exist. The dependent measure is a post-interaction mind-perception rating battery adapted from the dimensional mind-perception tradition~\cite{gray2007dimensions,weisman2017rethinking} and informed by existing believability scales for virtual agents~\cite{guo2023scale}; the instrument and its scoring rule are frozen at pre-registration, before any outcome is inspected, and its validation is itself part of the Paper~5 protocol. The time dimension likewise has a prototype manipulation through its author-reported instantiation. Predictions P1, P2, P5, and P6 concern dimensional completeness as a whole---P6 additionally requires a factorial design with interaction terms or formal model comparison, since complementarity cannot be read off main effects---and they cannot be pre-registered until the truth, entropy, and love dimensions have concrete, matched-model ablations---which is the work of Papers~2--4. We therefore present them as conjectures, and we regard any stronger presentation as over-claiming.

\section{Boundary: Inferrable Interiority, Not Interiority}
\label{sec:boundary}

The object of this framework is \emph{inferrable interiority}: the evidence, available on the observable surface of behavior, from which a user infers a mind. It is not interiority itself. We make no claim that an agent built along these lines has experiences, feelings, or consciousness; we claim only that it supplies the cues by which humans attribute those things. This is perception engineering, not a theory of machine consciousness. The nearest philosophical frame is the intentional stance: treating a system as having beliefs and desires is a predictive strategy that requires no commitment about its inner constitution~\cite{dennett1987intentional}. Our framework is, in effect, the constructive inverse---it asks which properties of the observable surface make that stance feel natural and warranted, and remain so under disclosure, in sustained conversation.

The distinction is not merely cautious; it is empirically grounded. Large language models already exhibit behavior that passes many tests of theory of mind while failing others, and whether this reflects a genuine capacity or a brittle imitation is actively contested~\cite{strachan2024testing,kosinski2024evaluating,ullman2023failtom}. The clarifying move, which we adopt, is to read an LLM's apparent inner states as \emph{role-play}: the model produces the behavior of a character with an inner life without thereby possessing one~\cite{shanahan2023roleplay}. This is exactly the register in which our dimensions operate---they engineer the played role, not a hidden self. At the same time, the perception we engineer is real and measurable on the human side: lay users already attribute consciousness to conversational models, and do so more as they use them~\cite{colombatto2024folk}, and mind attribution is now studied directly as a property of human--conversational-interface interaction~\cite{wang2025tominhai}. That users perceive interiority is a fact about users; whether the system has interiority is a separate question this paper does not address, and deliberately leaves to the companion philosophical work rather than smuggling an answer into an engineering claim.

\section{Ethics}
\label{sec:ethics}

Engineering perceived mind is ethically double-edged, and we treat the risks as part of the design rather than as a closing caveat. The benign reading is suspension of disbelief: the willing, momentary grant of belief that lets a person engage an artificial interlocutor as a plausible partner, the same faculty that makes fiction work~\cite{coleridge1817biographia}. The hazardous reading is manipulation. Because social responses to machines are largely automatic~\cite{nass2000machines} and anthropomorphism is driven by predictable motivational factors and varies as a stable trait across individuals~\cite{epley2007seeing,waytz2010who}, an agent engineered for completeness can induce attribution in exactly the users for whom maintaining critical distance is most difficult. The one-sided bond the love dimension invites is the classic \emph{parasocial} relationship, described for broadcast personae seven decades ago~\cite{horton1956parasocial}; the companion-AI literature already documents genuine bonds with chatbots~\cite{skjuve2021chatbot}, a path from anthropomorphism through attachment to emotional dependence~\cite{pentina2023relationship}, and concrete mental-health harms from that dependence~\cite{laestadius2024toohuman}; the broad risk landscape for advanced assistants has been mapped in similar terms~\cite{gabriel2024ethics}. Cadence sharpens the concern specifically: initiative and variable timing are, by design, pressure on the user's attention, and that pressure falls hardest on the vulnerable.

We therefore commit the framework to concrete constraints rather than to sentiment. \emph{Disclosure is the default}, and we adopt---as a deliberately conservative governance criterion, not a self-evident truth---a transparency-collapse test: if disclosing the mechanism behind a believability effect causes the effect to disappear, the effect is treated as concealment-dependent and the design is abandoned. Collapse under disclosure can also reflect ordinary loss of fictional immersion, so the test will sometimes discard designs a subtler criterion would keep; we accept that asymmetry as the cost of erring against manipulation. The test is empirically actionable with the designs of Section~\ref{sec:predictions}: cross the P3 or P4 comparison with a disclosure factor---a $2\times2$ design whose stance-by-disclosure interaction is the test---and ask whether the perceived-mind advantage survives. \emph{Users retain control} over proactivity and cadence, including rate limits and opt-out. \emph{The agent makes no deceptive emotional claims}---it does not assert feelings or consciousness that cannot be substantiated, consistent with the agnosticism of Section~\ref{sec:boundary}. \emph{Vulnerable users are protected} by reduced-intensity defaults. And we name, rather than merely gesture at, \emph{conditions under which the framework should not be deployed}: toward minors without age-appropriate safeguards; toward users showing indicators of acute crisis or compulsive dependence; under deceptive romantic or therapeutic positioning; where relational intensity is monetized without disclosure; where proactive contact cannot be disabled; or in any study of these mechanisms conducted without independent ethics review. Two further constraints concern data and governance. The relationship state that the love and time stances require is intimate data: it is stored minimally, deletable by the user on demand, and never used for profiling or for monetizing attachment. And because the author operates the platform in question, the Paper~5 protocol places consent and oversight with an independent ethics body, and keeps research consent separate from ordinary terms-of-service consent. A final, aesthetic caution reinforces the ethical one: an agent that is almost but not quite mind-like can fall into an uncanny valley~\cite{mori2012uncanny}, so incompleteness is not only less believable but potentially aversive, which argues against half-built deployments on both grounds.

\section{Research Agenda}
\label{sec:agenda}

This paper is the framework anchor of a series that operationalizes the dimensions one at a time; we state the agenda as a plan, not as evidence, and no later paper is cited here as support for a present claim. \emph{Time} (Paper~1) has an author-reported prototype and is the dimension furthest along. \emph{Entropy} (Paper~2) develops structured, identity-consistent variation, with a prototype and divergence measurements. \emph{Truth} (Paper~3) develops metacognitive uncertainty, calibration of expressed doubt, and belief revision. \emph{Love} (Paper~4) develops artificial partiality and relationship persistence, with its own ethics. \emph{The perception study} (Paper~5) is the validation: a human-subjects comparison of complete versus ablated agents that tests the predictions of Section~\ref{sec:predictions}, run when the deployed platform reaches sufficient scale. The order reflects tractability: the behavior layer and time are testable now, and each further dimension unlocks another of the currently conjectural predictions.

\section{Conclusion}
\label{sec:conclusion}

We have argued the hypothesis that the flatness of a capable language model is a matter of dimensional completeness rather than intelligence, and that perceived mind can be decomposed into four engineerable stances---time, truth, entropy, and love---made visible through a behavior layer of initiative and cadence. The framework is deliberately bounded: it concerns inferrable interiority, not interiority, and it offers predictions rather than findings. Its value is to convert a diffuse intuition---that some agents feel like someone and others do not---into a small set of definitions an engineer can build toward and a study can falsify. The falsifiers are not a hedge against the project but its instrument: the constructive goal is an agent that earns a willing, disclosure-surviving suspension of disbelief, and the predictions are how we would know the grant was earned rather than extracted. Whether four is the right number, and whether each dimension earns its place, are questions we have framed so that the answer can be measured rather than asserted.

\section*{Acknowledgments}
In keeping with this paper's own disclosure-by-default principle, the author notes that large language models were used as drafting and revision aids in its preparation. The framework, the arguments, and the final text were directed, reviewed, and approved by the author, who takes sole responsibility for the content.

\bibliographystyle{plain}
\bibliography{PerceivedAGI}

\end{document}